\documentclass[aps,pre,showpacs,twocolumn,amsmath]{revtex4-1}
\usepackage{graphicx}

\begin{document}

\title[]{Quantum critical behavior of the quantum Ising model on fractal lattices}
\author{Hangmo \surname{Yi}}
\email{hyi@ssu.ac.kr}
\affiliation{Department of Physics, Soongsil University, Seoul 156-743, Korea}
\affiliation{Institute for Integrative Basic Sciences, Soongsil University, Seoul 156-743, Korea}

\begin{abstract}
I study the properties of the quantum critical point of the transverse-field quantum Ising model on various fractal lattices such as the Sierpi\'nski carpet, Sierpi\'nski gasket, and Sierpi\'nski tetrahedron. Using a continuous-time quantum Monte Carlo simulation method and the finite-size scaling analysis, I identify the quantum critical point and investigate its scaling properties. Among others, I calculate the dynamic critical exponent and find that it is greater than one for all three structures. The fact that it deviates from one is a direct consequence of the fractal structures not being integer-dimensional regular lattices. Other critical exponents are also calculated. The exponents are different from those of the classical critical point, and satisfy the quantum scaling relation, thus confirming that I have indeed found the quantum critical point. I find that the Sierpi\'nski tetrahedron, of which the dimension is exactly two, belongs to a different universality class than that of the two-dimensional square lattice. I conclude that the critical exponents depend on more details of the structure than just the dimension and the symmetry.
\end{abstract}


\maketitle

\section{introduction}

The transverse-field quantum Ising model is one of the most widely used models for studying quantum effects on the magnetic order and critical phenomena in spin systems.\cite{sachdev1999booka,pfeuty1970a,porras2004a,grimmett2008a,friedenauer2008a}
In this model, quantum fluctuations are introduced by applying a magnetic field $\Delta$ perpendicular to the Ising spin direction.
Starting from the zero field limit, which corresponds to the original classical Ising model, one may investigate the effect of quantum fluctuations by controlling $\Delta$.
Recent theoretical studies\cite{yi2003a,yi2008a,yi2010a,yi2013a} of the ferromagnetic quantum Ising model based on various structures such as small-world networks, scale-free networks, and fractal lattices show that if one increases the transverse magnetic field, the ferromagnetic-paramagnetic phase transition temperature $T_c$ decreases monotonically.
However, the transverse field is found not to affect the critical exponents $\alpha,\ \beta,\ \gamma$ and $\nu$.
In addition, these studies all suggest that as the transverse magnetic field becomes strong enough, $T_c$ apparently vanishes at a critical field $\Delta_c$, but the limitation of the numerical method used in those works prohibited direct investigation of the zero temperature limit.

When $T=0$, the phase transition is controlled solely by $\Delta$ and is governed by the quantum critical point, which belongs, in general, to a different universality class from that of the classical counterparts away from $\Delta_c$.
One of the important characteristics of the quantum critical point is the dynamic critical exponent $z$.
It determines the relative scaling of space and time which leaves the action invariant in the vicinity of the quantum critical point, in the renormalization-group analysis.
In particular, it is known that $z=1$ for the transverse-field quantum Ising model in all integer-dimensional regular lattices.\cite{sachdev1999booka}
A recent study\cite{baek2011a} of the quantum Ising model on Watz-Strogatz small world networks also obtained results that are consistent with the above general rule for the integer-dimensional models in spite of the complex nature of the network structure.
It is in fact expected, because their model is in the mean-field limit and the upper critical dimension is an integer.

Fractal lattices are self-similar systems with many exotic physical properties.\cite{gouyet1996booka,mandelbrot1977booka} Since the dimension of a fractal is usually fractional, these structures have been a popular subject of research, especially as a tool to interpolate between integer-dimensional regular lattices. When dynamic objects are attached to the sites or vertices of such a system, one may study the statistical mechanics of a system in fractional dimensions.\cite{gefen1980a,gefen1981a} For example, many research papers have been devoted to the critical behavior of the {\em classical} Ising model on fractals.\cite{bonnier1988a,angles_d'auriac1986a,monceau1998a} More recently, fractal lattices have been also used in the {\em quantum} Ising model.\cite{yi2013a,yoshida2014a} 
Although the quantum model on fractals are a subject of only theoretical research and might not have direct practical implications at present, the rapid technological advancement will eventually drive the experimental system sizes to the quantum limit, and the quantum model may become more relevant in the practical applications, too.

In this paper, I will study the quantum critical point of the quantum Ising model on fractal lattices.
Especially, I will focus on the calculation of the dynamic critical exponent, because it is an interesting question whether $z$ is equal to one or not when the base structure is not an integer-dimensional regular lattice.
I will also compute other critical exponents and compare them with those of the classical critical point.

This paper is organized as follows.
In Sec.~\ref{sec:model}, I introduce the quantum Ising model and explain the method of quantum Monte Carlo simulation and finite-size scaling.
The method is applied to three fractal lattices and a two-dimensional square lattice in Sec.~\ref{sec:results}.
Finally I conclude with a summary and discussion in Sec.~\ref{sec:summary}.
The main results are summarized in Table~\ref{table:summary}.

\section{model and simulation method}
\label{sec:model}

The Hamiltonian of the quantum Ising model is given by
\begin{equation}
  H = -J \sum_{\left<ij\right>} \sigma^z_i \sigma^z_j + \Delta \sum_i \sigma^x_i
\end{equation}
where $\sigma^x_i$ and $\sigma^z_i$ are Pauli matrices representing the $x$ and $z$ components of the spin at site $i$, and the first summation runs only over nearest neighbor pairs.
We will consider only the ferromagnetic case($J>0$).
Apart from the temperature $T$, there are two important energy scales in this model: the ferromagnetic coupling constant $J$ and the transverse magnetic field $\Delta$.
For simplicity I will use the energy unit in which $J=1$ and the temperature unit in which $k_B=1$.
For $\Delta=0$, this model is identical to the ordinary classical Ising model and it is straightforward to obtain all energy eigenstates of the problem.
If $\Delta\neq 0$, however, the second term causes quantum fluctuations to the previous eigenstates because it does not commute with the first term.
As a result, it tends to destroy the ferromagnetic order that may have been resulted from the Ising exchange interaction.

This model is most easily analyzed using the Suzuki-Trotter decomposition method.\cite{suzuki1976a}
Writing the action as an integral in the imaginary time using the standard procedure, the temporal segments of each spin may be thought of as interacting via a nearest neighbor interaction in the time direction within each site.
Therefore, a quantum Ising model on a $D$-dimensional regular lattice may be mapped to a $(D+1)$-dimensional classical Ising model.
Since the imaginary time direction of the quantum model is simply one of the spatial directions of the classical counterpart, the dynamic critical exponent is simply given by $z=1$, as long as the original quantum system is an integer-dimensional regular lattice.
However, this argument does not apply to a fractal, because the spatial dimensions, being fractional, are no longer identical in nature to the temporal dimension.
The quantum Ising model on a fractal lattice has been studied in a previous paper\cite{yi2013a} in terms of the critical behavior at finite temperatures.
In the current work, I will focus on the quantum critical point at $T=0$.

We cannot directly access the zero temperature limit in Monte Carlo simulations, because the length in the time direction is inversely proportional to $T$ and becomes infinity.
Instead, I will rely on the finite size scaling method.
In order to identify the critical transverse field $\Delta_c$, I will use the fourth-order Binder cumulant\cite{binder1981a}
\begin{equation}
  U = 1 - \frac{\left< m^4\right>}{3\left< m^2\right>^2}.
\end{equation}
Here, $m$ is the magnetization per spin and $\left<\cdots\right>$ denotes the thermal average.
In the vicinity of the quantum critical point, this quantity obeys a finite-size scaling form
\begin{equation}
  U(T,\Delta,L) = \tilde{U}\left( (\Delta-\Delta_c)L^{1/\nu},TL^z \right),
  \label{eq:scaling_binder}
\end{equation}
where $L$ is the system size and $\nu$ is the critical exponent for the correlation length.
If $\Delta=\Delta_c$, the above quantity depends only on $TL^z$.
Therefore, we can identify $\Delta_c$ by demanding that the maximum of $\tilde{U}$ as a function of $T$ should not depend on $L$.
Then $z$ is obtained from the condition that $\tilde{U}(0,TL^z)$ for different system sizes should all collapse onto a single curve within the scaling regime.
Other critical exponents may also be found using the following scaling functions for the magnetization $m$ and the magnetic susceptibility $\chi$.
\begin{eqnarray}
  \label{eq:m}
  m(T,\Delta,L) & = & L^{-\beta/\nu}\tilde{m}\left( (\Delta-\Delta_c)L^{1/\nu},TL^z \right), \\
  \label{eq:chi}
  \chi(T,\Delta,L) & = & L^{\gamma/\nu}\tilde{\chi}\left( (\Delta-\Delta_c)L^{1/\nu},TL^z \right).
\end{eqnarray}

I developed a quantum Monte Carlo simulation program using the Swendesen-Wang cluster algorithm.\cite{swendsen1987a}
In order to handle the imaginary time dimension, I adopted a continuous-time method,\cite{rieger1999a} the details of which is briefly outlined below.
The world line of each spin, which is considered a continuous object, may be divided into many segments with any arbitrary length and spin value(up or down).
At the start of every Monte Carlo step, new cutting lines are inserted at random positions of the world lines, further dividing the segments via a Poisson process.
Then any two segments of neighboring spins are allowed to be connected and belong to the same cluster, according to the probability determined by the time overlap between them.
Finally, each cluster made up of connected segments are randomly assigned a new spin value.
After removing redundant cuts between adjacent segments with the same spin value, the Monte Carlo step is concluded.
This procedure is repeated until desired accuracy is obtained for the physical quantity of interest.
This method is especially efficient near a critical point, where all different lengths of segments are present.

\begin{figure}[ht]
\resizebox{0.6\columnwidth}{!}{%
  \includegraphics{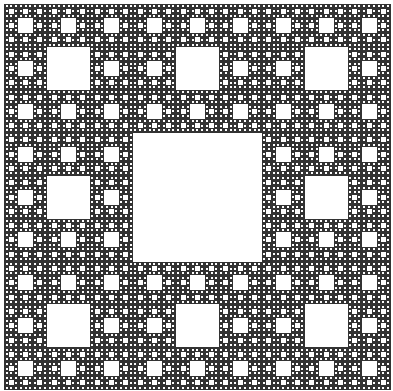}
}
\caption{
  The Sierpi\'nski carpet constructed by the method explained in the text. The recursion process has been repeated five times for this example.
}
\label{fig:sierpinski_carpet}
\end{figure}

\begin{figure*}
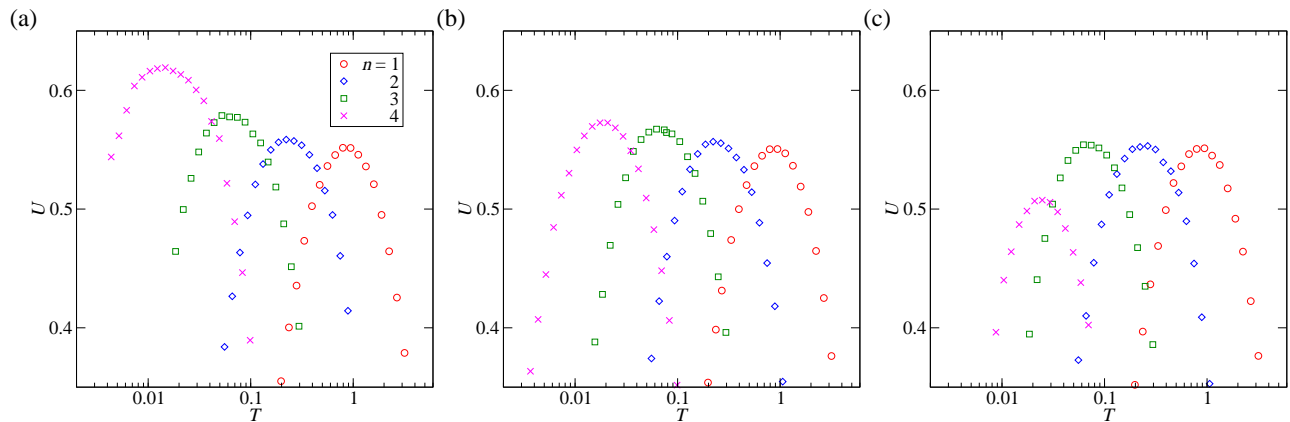

\resizebox{0.95\textwidth}{!}{%
  \includegraphics{fig-binder2.395.eps}
  \includegraphics{fig-binder2.3975.eps}
  \includegraphics{fig-binder2.4.eps}
}
\caption{
  (Color online)
  Binder cumulant $U$ as a function of temperature $T$ for (a) $\Delta=2.395$, (b) 2.3975, and (c) 2.4. The system size is $L=3^n$, where $n$ is the number of recursions. The errorbars are smaller than the symbols.
}
\label{fig:binder}
\end{figure*}

\begin{figure*}
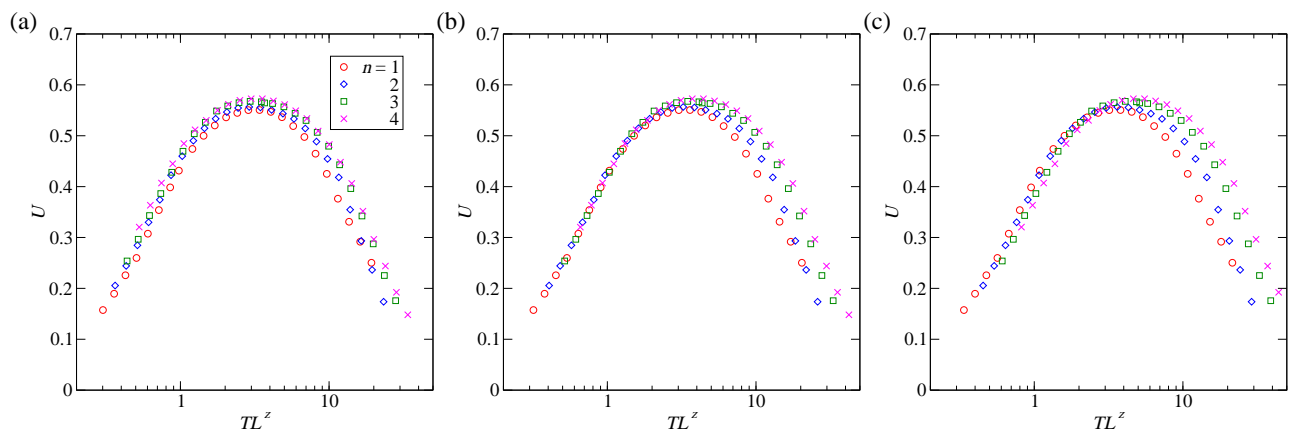

\resizebox{0.95\textwidth}{!}{%
  \includegraphics{fig-scaling1.17.eps}
  \includegraphics{fig-scaling1.22.eps}
  \includegraphics{fig-scaling1.27.eps}
}
\caption{
  (Color online)
  Scaling function $\tilde{U}(0,TL^z)$ for (a) $z=1.17$, (b) 1.22, and (c) 1.27. The errorbars are smaller than the symbols.
}
\label{fig:scaling_T}
\end{figure*}

\section{analysis and results}
\label{sec:results}

\subsection{Sierpi\'nski carpet}

The first fractal lattice I will use in this paper is the Sierpi\'nski carpet, which is constructed in the following way.
First, a two dimensional square lattice is divided into a $3\times 3$ array of equal-size square regions and then the region in the middle is removed.
Each of the remaining regions is again divided into a $3\times 3$ array of smaller square regions, and this process is repeated recursively.
Figure \ref{fig:sierpinski_carpet} shows the result of this process after five recursions.
After $n$ recursions, the total number of sites is $8^n$ and the length of each side $3^n$, hence the Hausdorff dimension of this structure is $d_\mathrm{H} = \log 8/\log 3\approx 1.893$.
In our quantum Ising model, each site of the Sierpi\'nski carpet is occupied by an Ising spin. We will assume a periodic boundary condition, where the sites in the rightmost column are connected to the ones in the leftmost column. The top and bottom rows are connected likewise.\cite{boundary}

The Binder cumulant calculated from the numerical simulations is shown in Fig.~\ref{fig:binder}.
From the condition of the maximum value being independent of $L$, we find that $\Delta_c=2.3975\pm 0.0025$.
This is consistent with the approximate result obtained from the finite temperature analysis in Ref.~\onlinecite{yi2013a}, where it was speculated that $\Delta_c\approx 2.4$.
The universal maximum value of the Binder cumulant may be simply read off from the data and is given by $U^*_\mathrm{max}=0.57\pm 0.02$.

The scaling function $\tilde{U}$ at $\Delta_c$ as a function of the single parameter $TL^z$ is shown in Fig.~\ref{fig:scaling_T}.
Focusing on the data close to the quantum critical point at $T=0$, it is obvious that the middle plot [Fig.~\ref{fig:scaling_T}(b)] shows the best collapse and we obtain $z=1.22\pm 0.05$.
Since this is not equal to one, we may conclude that the dynamic critical exponent in our fractal lattice system is not the same as in integer-dimensional regular lattices.

\begin{figure*}
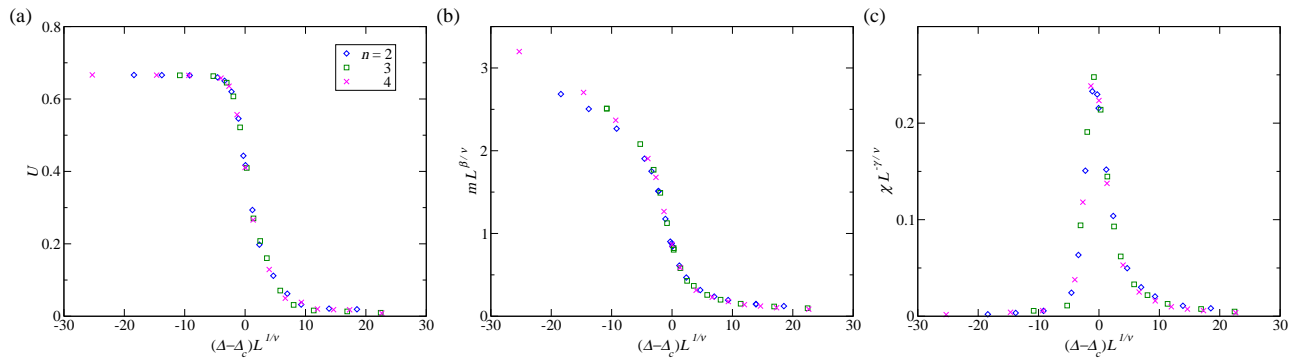

\resizebox{0.95\textwidth}{!}{%
  \includegraphics{fig-scaling-binder.eps}
  \includegraphics{fig-scaling-mag.eps}
  \includegraphics{fig-scaling-sus.eps}
}
\caption{
  (Color online)
  Scaling functions (a) $\tilde{U}$, (b) $\tilde{m}$, and (c) $\tilde{\chi}$. I used $z=1.22$ while keeping $TL^z=1$ for all system sizes. The critical exponents used here are $\nu=0.70$, $\beta=0.37$, and $\gamma=1.45$. The errorbars are smaller than the symbols.
}
\label{fig:scaling_delta}
\end{figure*}

The critical exponent $\nu$ may be obtained using Eq.~(\ref{eq:scaling_binder}). If we keep $TL^z$ constant, $\tilde{U}$ must be a function of a single parameter $(\Delta-\Delta_c)L^{1/\nu}$.
Figure \ref{fig:scaling_delta}(a) shows that the scaling curves collapse nicely into one single curve, from which I estimate $\nu=0.70\pm0.05$.
The precision in this result is estimated by tuning $\nu$ around the best fitting value and finding the limit where it is acceptable to regard all curves with different system sizes to collapse into one.
Using Eqs.~(\ref{eq:m}) and (\ref{eq:chi}) with Figs.~\ref{fig:scaling_delta}(b) and (c), I can also obtain $\beta=0.37\pm0.03$ and $\gamma=1.45\pm0.08$.
Now we may check whether these exponents satisfy the scaling relations.
Especially, we find that
\begin{equation}
  d_\mathrm{scaling} = \frac{2\beta + \gamma}{\nu} = 3.1\pm 0.3.
\end{equation}
The spatial and the temporal size of the system in the vicinity of the quantum critical point scale as $L^{d_\mathrm{H}}$ and $L^z$, respectively. Therefore, the effective dimension of the quantum critical system in the scaling limit is given by
\begin{equation}
  d_\mathrm{eff} = z + d_\mathrm{H} = (1.22\pm0.05) + \frac{\log 8}{\log 3} = 3.11\pm 0.05.
\end{equation}
The agreement between $d_\mathrm{scaling}$ and $d_\mathrm{eff}$ is exceptional, which confirms that we have indeed found a quantum critical point.

Now let us compare these critical exponents to those of the classical critical point at finite $T_c$. According to Ref.~\onlinecite{yi2013a}, they are given by $\nu=1.62\pm 0.05$, $\beta= 0.13\pm 0.01$, and $\gamma= 2.85\pm 0.05$.
These values are very different from those of the quantum case discussed above, which proves that the quantum and the classical critical points belong to different universality classes.
It is worth mentioning that the scaling relation is also well satisfied for the classical critical point.
\begin{gather}
  d_\mathrm{scaling}^\mathrm{classical} = \frac{2\beta + \gamma}{\nu} = 1.9\pm 0.2 \\
  d_\mathrm{eff}^\mathrm{classical} = \frac{\log 8}{\log 3} \approx 1.893.
\end{gather}

Before we move on, I need to comment on a contradicting result in Ref.~\onlinecite{yoshida2014a}, in which the authors claim that the quantum phase transition for the quantum Ising model on the Sierpinski carpet is a weak first-order transition. In the above analysis, however, all evidence including the scaling relation indicates that it is a second order phase transition.\cite{yoshida_z=1}

\subsection{Sierpi\'nski gasket}

\begin{figure}[ht]
  (a) \vtop{\null\hbox{\includegraphics[scale=0.2]{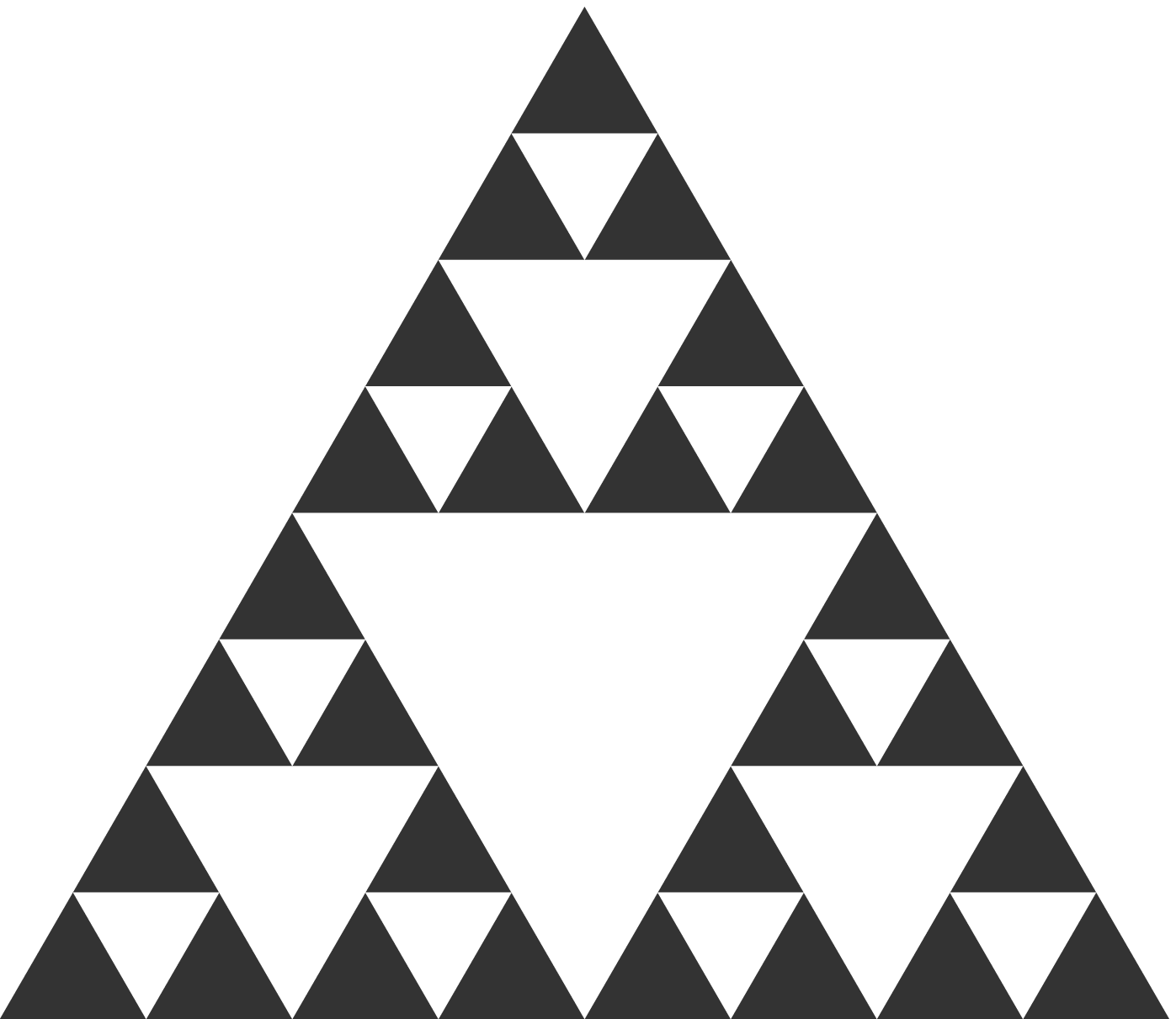}}}
  \hspace{2em}
  (b) \vtop{\null\hbox{\includegraphics[scale=0.5]{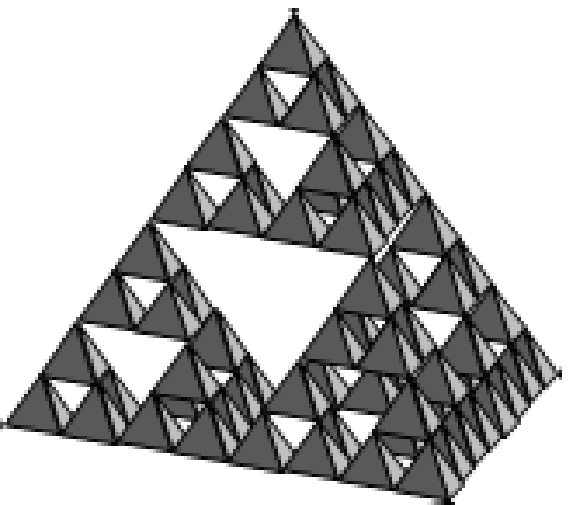}}}
\caption{
  (a) A Sierpi\'nski gasket can be constructed in a similar manner as with the Sierpi\'nski carpet, but using triangles instead of squares. (b) A Sierpi\'nski tetrahedron is a three-dimensional analogue of a Sierpi\'nski gasket.
}
\label{fig:sierpinski-gasket-tetrahedron}
\end{figure}

I will now consider the Sierpi\'nski gasket such as shown in Fig.~\ref{fig:sierpinski-gasket-tetrahedron}(a), of which the Hausdorff dimension is given by $d_\mathrm{H} = \log 3/\log 2 \approx 1.585$. This may be constructed in a very similar manner as we did in Sec.~\ref{sec:results}.A. However, this structure has a finite ramification number, which means any arbitrarily large sublattice may be separated by cutting a finite number of bonds. The Sierpi\'nski carpet, on the other hand, is infinitely ramified. This seemingly subtle difference affects the phase transition rather significantly.\cite{havlin1984a,hao1987a} Most important of all, the (classical) Ising model on a fractal lattice with a finite ramification order cannot have a phase transition at finite temperatures.\cite{gefen1980a,carmona1998a} However, this argument does not apply to the {\em quantum} Ising model, because when it is mapped to a classical model with one more dimension following the Suzuki-Trotter method, the resulting system is infinitely ramified.

\begin{figure*}
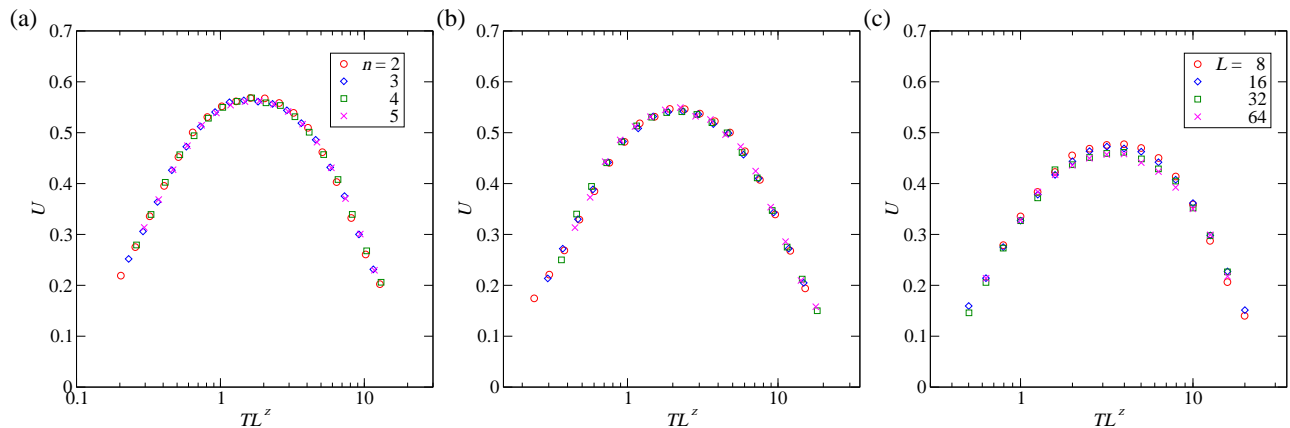

\resizebox{0.95\textwidth}{!}{%
  \includegraphics{fig-scaling-gasket1.18.eps}
  \includegraphics{fig-scaling-tetrahedron1.3.eps}
  \includegraphics{fig-scaling-square1.eps}
}
\caption{
  (Color online)
  The scaling plots of the Binder cumulant for (a) the Sierpi\'nski gasket ($z=1.18$ and $\Delta=1.865$), (b) the Sierpi\'nski tetrahedron ($z=1.30$ and $\Delta=2.707$), and (c) the two-dimensional square lattice ($z=1,\ \Delta=3.045$).
}
\label{fig:scaling-others}
\end{figure*}

We will assign a spin to each triangle, and will assume that any two triangles are nearest neighbors if they share a common vertex. I have chosen a special boundary condition that allows fast convergence in the simulation.\cite{yoshida2014a} Specifically, I made two copies of the same Sierpi\'nski gasket and connected each of the three corners of one copy to the corresponding corner of the other.

Figure~\ref{fig:scaling-others}(a) shows the scaling of the Binder cumulant near the critical point $\Delta=1.865\pm0.005$. The dynamic critical exponent and the universal maximum of the Binder cumulant are estimated to be $z = 1.18 \pm 0.05$ and $U^*_\mathrm{max} = 0.56\pm0.01$. Other critical exponents are given by $\nu=0.66\pm 0.05$, $\beta = 0.19\pm 0.02$, and $\gamma = 1.45\pm 0.05$. The effective dimension obtained from the scaling relation is given by $d_\mathrm{scaling} = (2\beta+\gamma)/\nu = 2.8\pm 0.2$. This agrees very well with the effective dimension $d_\mathrm{eff} = z + d_\mathrm{H} = 2.77 \pm 0.05$. This confirms that the scaling hypothesis works well for the Sierpi\'nski gasket.

\subsection{Sierpi\'nski tetrahedron}

Now let us examine the model on the Sierpi\'nski tetrahedron.[Fig.~\ref{fig:sierpinski-gasket-tetrahedron}(b)] This structure is especially interesting because its Hausdorff dimension is exactly two. ($d_\mathrm{H} = \log 4/\log 2 = 2$) Although its dimension is the same as that of a two-dimensional regular lattice, their structures are fundamentally different. Therefore, it may be used to test whether the universality class depends only on the dimension and the symmetry, as is often stated.

For the Sierpi\'nski tetrahedron, I obtained $\Delta_c = 2.707\pm0.002$ and $U^*_\mathrm{max}=0.54\pm0.01$. The critical exponents are estimated to be $z=1.30\pm 0.05$, $\nu=0.62\pm 0.05$, $\beta=0.25\pm 0.02$, and $\gamma=1.55\pm 0.05$. It is quite interesting that $z\neq 1$ even though the Hausdorff dimension of this system is an integer. Once again, we can use the scaling relation to get $d_\mathrm{scaling} = (2\beta+\gamma)/\nu = 3.3\pm 0.2$, which is the same as the effective dimension $d_\mathrm{eff} = z + d_\mathrm{H} = 3.30\pm 0.05$. One may compare these exponents with those of the quantum Ising model on a two-dimensional lattice. Since a $D$-dimensional quantum model is mapped to a $(D+1)$-dimensional classical model, the quantum critical exponents of a two-dimensional regular lattice are the same the classical critical exponents of a three-dimensional regular lattice: $\nu=0.63012(16)$, $\beta=0.32653(10)$, and $\gamma=1.2373(2)$.\cite{campostrini2002a} Note the exponents do not agree between the two systems. Therefore, we conclude that they belong to different universality classes, even though they have the same dimension and symmetry.

Finally, I applied the analysis to the two-dimensional square lattice in order to check the validity of the analysis technique used here. The scaling plot of the Binder cumulant is shown in Fig.~\ref{fig:scaling-others}(c). I obtained $\Delta_c = 3.045\pm0.003$ and $U^*_\mathrm{max} = 0.47\pm0.01$. Most importantly, I estimated $z=1.00\pm 0.05$, which is consistent with the expectation that $z=1$ for an integer-dimensional regular lattice. The other exponents are estimated to be $\nu=0.63\pm0.05$, $\beta=0.33\pm0.03$, and $\gamma=1.24\pm 0.05$. They are all in good agreement with the known values, providing evidence that the analysis technique is legitimate.

\section{summary and discussion}
\label{sec:summary}

\begin{table*}[bth]
\centering
\begin{math}
\begin{array}{c||c|c|c|c|c|c|c|c|c||c|c|c}
\hline\hline
$lattice$ & d_\mathrm{H} & \Delta_c & U^*\mathrm{max} & z & \nu & \beta & \gamma & d_\mathrm{scaling} & d_\mathrm{eff} & \nu_\mathrm{\epsilon} & \beta_\mathrm{\epsilon} & \gamma_\mathrm{\epsilon} \\
\hline
\hspace*{-1em} $\parbox[c]{6em}{Sierpi\'nski carpet}$ & 1.893 & 2.3975(25) & 0.57(2) & 1.22(5) & 0.70(5) & 0.37(3) & 1.45(8) & 3.1(3) & 3.11(5) & 0.61(1) & 0.35(1) & 1.21(1) \\
\hspace*{-1em} $\parbox[c]{6em}{Sierpi\'nski gasket}$ & 1.585 & 1.865(2) & 0.56(1) & 1.18(5) & 0.66(5) & 0.19(2) & 1.45(5) & 2.8(2) & 2.77(5) & 0.68(1) & 0.28(1) & 1.32(1) \\
\hspace*{-1em} $\parbox[c]{6em}{Sierpi\'nski tetrahedron}$ & 2 & 2.707(2) & 0.54(1) & 1.30(5) & 0.62(5) & 0.25(2) & 1.55(5) & 3.3(2) & 3.30(5) & 0.58(1) & 0.38(1) & 1.15(1) \\
\hspace*{-1em} $\parbox[c]{6em}{2D square lattice}$ & 2 & 3.045(3) & 0.47(1) & 1.00(5) & 0.63(5) & 0.33(3) & 1.24(5) & 3.0(2) & 3.00(5) & 0.6310(15) & 0.3270(15) & 1.2390(25) \\
\hline\hline
\end{array}
\end{math}
\caption{Summary of the results from the numerical analysis for various lattice structures. $d_\mathrm{H}$ is the Hausdorff dimension, $d_\mathrm{scaling}=(2\beta+\gamma)/\nu$ is the dimension calculated from the scaling relation, and $d_\mathrm{eff}=d_\mathrm{H}+z$ is the effective dimension of the quantum critical point in the scaling limit. $\nu_\mathrm{\epsilon}$, $\beta_\mathrm{\epsilon}$, $\gamma_\mathrm{\epsilon}$ are linearly interpolated values from the results of an $\epsilon$-expansion study in Ref.~\onlinecite{le_guillou1987a}, using $d_\mathrm{eff}$. The numbers in parentheses denote the uncertainty in the last digits.}
\label{table:summary}
\end{table*}

In summary, I have studied the quantum critical point of the transverse-field quantum Ising model on several fractal lattices.
Using a continuous-time quantum Monte Carlo simulations and a finite-size scaling method on the Sierpi\'nski carpet, gasket, and tetrahedron, I identified the quantum critical point for each structure and calculated its various physical quantities including the critical transverse field, the universal maximum of the Binder cumulant, and the critical exponents.
The main results of this paper are summarized in Table~\ref{table:summary}.
For all three structures, $z$ is not equal to one, which is a direct manifestation of the system not being an integer-dimensional regular lattice. I find that the scaling relation $(2\beta + \gamma)/\nu=d$ is well satisfied in all cases, if we assume that $d=d_\mathrm{H}+z$ where $d_\mathrm{H}$ is the Hausdorff dimension of the fractal lattice.
For the Sierpi\'nski carpet, the quantum critical exponents were explicitly shown to be different from those of the classical critical point.
All these results confirm that I have indeed found the quantum critical point.

In Ref.~\onlinecite{le_guillou1987a}, Le Guillou and Zinn-Justin have used an $\epsilon$-expansion technique to compute various critical exponents for fractional dimensions between 1 and 4. (See Table~\ref{table:summary})
Note that the valued do not agree well with my results for all three fractal lattices.
This indicates that the critical exponents are {\em not} uniquely determined by the dimension and the symmetry alone.
It is even more apparent for the case of the Ising model on the Sierpi\'nski tetrahedron, which has the same dimension as a two-dimensional regular lattice, but does not belong to the same universality class.

\acknowledgments
This research was supported by the Basic Science Research Program through the National Research Foundation of Korea funded by the Ministry of Science, ICT and Future Planning (NRF-2012R1A1A2041460).

\bibliography{journal,stat-phys,cond-mat,mine,etc,sierpinski-quantum}

\begin{thebibliography}{29}
\bibitem [{yos()}]{yoshida_z=1}%
  \BibitemOpen
  \href@noop {} {}\bibinfo {note} {In Ref.~\onlinecite{yoshida2014a}, the
  authors made an assumption that $z=1$, which may be the reason for the
  difference.}\BibitemShut {Stop}%
\end{thebibliography}

\end{document}